\documentclass{elsart}
\usepackage{epsfig}

\def\gsim{\;\raise0.3ex%
\hbox{$>$\kern-0.75em\raise-1.1ex\hbox{$\sim$}}\;}
\def\lsim{\;\raise0.3ex%
\hbox{$<$\kern-0.75em\raise-1.1ex\hbox{$\sim$}}\;}

\begin{document}
\begin{frontmatter}

\hbox to\hsize{\hfill DSF 26/2004}
\hbox to\hsize{\hfill MPP-2004-121}

\title{Diffuse Cosmic Neutrino Background from Population III Stars}
\author[Napoli]{F.~Iocco},
\author[Napoli]{G.~Mangano},
\author[Napoli]{G.~Miele},
\author[Munich]{G.~G.~Raffelt},
\author[Munich]{P.~D.~Serpico}

\address[Napoli]{Dipartimento di Scienze Fisiche, Universit\`{a} di
Napoli ``Federico II''\break
and INFN, Sezione di Napoli\break
Complesso Universitario di Monte Sant'Angelo,
Via Cintia, 80126 Napoli, Italy}

\address[Munich]{Max-Planck-Institut f\"{u}r Physik
(Werner-Heisenberg-Institut)\break
F\"{o}hringer Ring 6, 80805 M\"{u}nchen, Germany}

\begin{abstract}
  We study the expected diffuse cosmic neutrino flux produced by
  Population III (PopIII) stars during their nuclear burning phases as
  well as from their final stages of evolution (core
  collapse). Assuming a fraction $f_{\rm III}=10^{-3}$ of all baryons
  forms PopIII stars, our flux estimate is comparable to
  the diffuse neutrino flux produced by the ordinary stars and by the
  ordinary core-collapse supernovae in the universe, i.e.\ of order
  1--$10~{\rm cm}^{-2}~{\rm s}^{-1}$.  Due to the large cosmic
  redshift, however, the typical energies are in the MeV and sub-MeV
  range where the solar and geophysical neutrino fluxes are much
  larger. A direct detection of the diffuse cosmic flux is out of the
  question with presently known experimental techniques.
\end{abstract}

\end{frontmatter}

%%%%%%%%%%%%%%%%%%%%%%%%%%%%%%%%%%%%%%%%%%%%%%%%%%%%%%%%%%%%%%%%%%%%%%
\section{Introduction}
%%%%%%%%%%%%%%%%%%%%%%%%%%%%%%%%%%%%%%%%%%%%%%%%%%%%%%%%%%%%%%%%%%%%%%

A pregalactic generation of stars known as Population III (PopIII) is
thought to have formed from the pristine metal-free gas left from
primordial nucleosynthesis.  The peculiar characteristics of this
hypothetical class of objects make it a favored candidate for the
engine driving the cosmic reionization event, put in the range
$11<z_r<30$ at 95\%~C.L.\ by the optical-depth measurements of WMAP
\cite{Kogut:2003et,Spergel:2003cb}.  The high-mass peak of the PopIII
initial mass function (IMF) \cite{Nakamura:2000ez,Abel:2000tu} should
produce both pair-instability supernovae if the stellar mass is
between 140~M$_{\odot}$ and 260~M$_{\odot}$ and black holes for both
smaller and larger masses~\cite{Heger:2001np,FWH00}.  Moreover, their
strong UV emission would reionize the surrounding space.

Unfortunately, the short evolutionary timescales of such massive stars
of about $10^{6}$~years and their presumed formation sites in
minihalos make them inaccessible to direct observations.  The massive
peak of the predicted bimodal IMF is probably the only part which
effectively contributes to the stellar history because both the
chemical and radiation feedback of these massive stars prevent the
formation of a low-mass population that otherwise could have survived
until now.  Therefore, bounds on their existence and properties could
be obtained indirectly by reionization data~\cite{reion1}, infrared
background analyses~\cite{Magliocchetti:2003hr}, or nucleosynthesis
yields~\cite{Daigne:2004ga,Heger:2001cd}.  Further details on this
issue can be found, for example, in Ref.~\cite{Bromm:2003vv}.  At
present, none of these observables provides clear evidence for the
existence of PopIII stars, nor do they rule out their occurrence in
the early universe.

Interestingly, an important fraction of the stars belonging to this
generation may experience a final fate with a huge emission of
neutrinos~\cite{FWH00}. Moreover, independently from their final fate,
these stars produce a considerable amount of thermonuclear neutrinos
during their main sequence evolution through the pp chains and
particularly the CNO cycle. In addition, they produce thermal
neutrinos in their late burning phases (C-Ne-O-Si).  The aim of this
paper is to estimate the diffuse cosmic neutrino flux originating from
these stars and to check whether there are any chances for its direct
detection with present experimental technology.

Our study has two main motivations. First, the lack of any direct
evidence of PopIII stars makes the clean signature represented by
their neutrino fluxes an appealing but challenging task.  Secondly, in
the era of precision neutrino observations it is mandatory to explore
and estimate all possible backgrounds for these experiments.  Recent
efforts have been made to evaluate the cosmic and stellar neutrino
background in the range of 0.1--few~MeV \cite{Porciani:2003zq} as well
as the neutrino and antineutrino solar flux in the very low energy
range~\cite{Haxton:2000xb}.  Therefore, it is worth looking for other
possible sources to properly estimate the total flux in these energy
ranges, and indeed we will show that PopIII are likely to represent a
significant cosmic contribution, though of limited interest for direct
experimental searches.

We begin in Sec.~\ref{sec:model} with a general formulation of the
expression for the diffuse cosmic neutrino flux from the PopIII
sources. In Secs.~\ref{sec:thermonuclear}, \ref{sec:thermal},
and~\ref{sec:corecollapse} we estimate the contributions from
thermonuclear reactions, thermal plasma processes, and the final
core-collapse phase. In Sec.~\ref{sec:conclusions} we summarize and
discuss our results.

\newpage

%%%%%%%%%%%%%%%%%%%%%%%%%%%%%%%%%%%%%%%%%%%%%%%%%%%%%%%%%%%%%%%%%%%%%%
\section{Schematic model for neutrino flux predictions}
%%%%%%%%%%%%%%%%%%%%%%%%%%%%%%%%%%%%%%%%%%%%%%%%%%%%%%%%%%%%%%%%%%%%%%

\label{sec:model}

The neutrino flux emitted by PopIII stars consists of several
contributions, covering different energy ranges.

(i) The neutrino flux from the hydrogen burning phase that would
contribute in the energy range 0.01--0.1~MeV because of cosmological
redshift effect.

(ii) The neutrinos and anti-neutrinos of all flavors produced in the
advanced burning stages by purely leptonic processes
\cite{Haft:1993jt,Esposito:2001if,Esposito:2003wv}.  This contribution
depends rather critically on the star's chemical composition,
temperature and density profile as well as the timescales of the
relevant evolutionary phases.

(iii) The huge emission of neutrinos and anti-neutrinos of all flavors
produced during the core collapse. The expected energy range is
0.1--1~MeV.

(iv) The neutrinos and anti-neutrinos of very high energy ($E>1$~TeV)
of a possible GRB phase~\cite{Schneider:2002sy}. After the collapse to
a black hole, an accretion disk could form and hadronic photo-meson
production would occur in strongly collimated and relativistic jets.

In the following we will describe in some detail the expected fluxes
for the first three contributions and compare them with other known
sources populating these energy ranges.  Since a detailed description
of the PopIII epoch is still lacking, especially as far as their
initial and final evolutionary stages are concerned, we will only give
order-of-magnitude estimates and consider a schematic model.  The
neutrinos from a possible GRB phase, though probably constituting the
most likely PopIII contribution to be detected, are also the most
model-dependent, and will not be further discussed here.  It is
however encouraging that the new AMANDA limits already significantly
constrain the parameter space adopted in Ref.~\cite{Schneider:2002sy},
meaning that a possible detection of these neutrinos is already within
the capabilities of present neutrino telescopes~\cite{Meszaros}.

Denoting by $\nu_\alpha$ a generic neutrino or anti-neutrino flavor
state, the differential flux from a production mechanism $j$ expected
at Earth is
\begin{equation}\label{flux}
\frac{dF^\alpha_j}{dE}=c \int_{z_{f}}^{z_{i}} \frac{dz}{H(z)}
R_{\rm III}(z) \int \frac{dM}{M}\,
\frac{dn(M)}{d\ln M} \,N^\alpha_j \,
\frac{dP^\alpha_j(M,E(1+z))}{dE}\,,
\end{equation}
where $R_{\rm III}(z)$ is the average-mass star-formation rate per
unit comoving volume, $dn(M)/d (\ln M)$ is the IMF, while $z_i$ and
$z_f$ denote the initial and final redshifts of the PopIII epoch.
Further, $N^\alpha_j$ is the total number of $\nu_\alpha$ emitted by a
star with mass $M$ due to the $j$-th mechanism (thermal, nuclear, core
collapse), integrated over the entire stellar lifetime and
$dP^\alpha_j(M,E(1+z))/dE$ is the corresponding emission probability
per unit of initial unredshifted energy $E(1+z)$.  Finally, $H(z)$ is
the Hubble parameter
\begin{equation}\label{hpar}
H(z)= H_0 \left[\Omega_{\rm M}
  (1+z)^3+ \Omega_\Lambda+
  (1-\Omega_{\rm M}-\Omega_\Lambda)(1+z)^2 \right]^{1/2}\,,
\end{equation}
where we have ignored the contribution of radiation to $\Omega$
because it is negligible for the relevant redshift range.  In the
following we will consider a flat cosmological model with $\Omega_{\rm
M}=0.3$ and $\Omega_\Lambda=0.7$.

The PopIII neutrino flux is quite sensitive to the assumed IMF.
Recent simulations of the star-formation phenomenon in a metal free
environment provide an IMF peaked at masses $M \sim 100$--$300\,{\rm
M}_\odot$ \cite{Abel:2000tu}.  Unless otherwise stated, we will
consider a simple model for the IMF, assuming that it is a peaked
function at $M_{\rm III}=300 {\rm M}_\odot$.  This is an optimistic
scenario, giving the largest neutrino flux from the dominating
core-collapse mechanism. Detailed modeling for this mass value has
been performed in Ref.~\cite{FWH00}.

The number of baryons forming a PopIII star of mass $M_{\rm III}$ is
$M_{\rm III}/m_N$ with $m_N$ the nucleon mass.  Denoting by $f_{\rm
  III}$ the total time-integrated fraction of baryonic matter which is
converted into PopIII stars we can re-cast Eq.~(\ref{flux}) as
\begin{equation}\label{flux2}
\frac{dF^\alpha_j}{dE}=c\, f_{\rm III}\,n_\gamma\,\eta
\,\frac{m_N}{M_{\rm III}}\,\int_{z_{f}}^{z_{i}} dz \, \rho(z)\,
(1+z)\,N^\alpha_j \, \frac{dP^\alpha_j(M_{\rm III},E(1+z))}{dE}\,,
\end{equation}
where $n_\gamma=410~{\rm cm}^{-3}$ is the number density of cosmic
microwave photons and $\eta=(6.3 \pm 0.3)\times 10^{-10}$ is the
cosmic baryon-to-photon ratio \cite{Spergel:2003cb}. Moreover,
\begin{equation}\label{rho}
 \rho(z) = \frac{R_{\rm III}(z)}{H(z)(1+z)}
 \left(\int_{z_f}^{z_i} dz\, \frac{R_{\rm III}(z)}{H(z)(1+z)}
  \right)^{-1}
\end{equation}
is a normalized star-formation rate per unit redshift. The functional
form of $\rho(z)$ enters mainly in determining the energy range where
the neutrino flux from PopIII is expected in view of the cosmic
redshift.

The value of $f_{\rm III}$ is probably the largest source of
uncertainty.  The existing estimates suggest a range
$10^{-4}$--$10^{-2}$ \cite{Abel:2000tu}. We will use
\begin{equation}
f_{\rm III}=10^{-3}
\end{equation}
in all of our estimates.  As $f_{\rm III}$ enters the neutrino flux
linearly, our results scale trivially for different assumed values of
this parameter.

To illustrate the sensitivity of our results on $\rho(z)$ we will
consider three generic cases.  The first is $\rho(z)$ peaked at some
definite value $\bar{z}$, i.e.\ $\rho(z) = \delta(z-\bar{z})$.  The
second is a constant star-formation rate as a function of redshift,
$\rho(z) = (z_i-z_f)^{-1}$.  Finally, we consider a constant star
formation rate as a function of time, $\rho(z) \propto
[H(z)(1+z)]^{-1}$.

%%%%%%%%%%%%%%%%%%%%%%%%%%%%%%%%%%%%%%%%%%%%%%%%%%%%%%%%%%%%%%%%%%%%%%
\section{Thermonuclear neutrinos}
%%%%%%%%%%%%%%%%%%%%%%%%%%%%%%%%%%%%%%%%%%%%%%%%%%%%%%%%%%%%%%%%%%%%%%

\label{sec:thermonuclear}

We begin with an estimate of the neutrino flux from the
hydrogen-burning phase.  Since the initial hydrogen mass fraction is
approximately 75\%, one would conclude that the upper limit to the
number of $\nu_e$ is given by $0.5\times 0.75\,M_{\rm III}/m_N\approx
0.38\, M_{\rm III}/m_N$.  By only considering the helium core masses
given in~Ref.~\cite{FWH00}, one can deduce that at least a fraction
$105/300=35\%$ ($M_{\rm III}=300\,{\rm M}_\odot$) or $67.5/250=27\%$
($M_{\rm III}=250\,{\rm M}_\odot$) of the initial stellar mass has
been converted into helium.  However, also the external layers of the
star are helium and metal enriched~\cite{FWH00}.  Therefore, a
reasonable estimate for the overall number of thermonuclear neutrinos
emitted per star is
\begin{equation}
N_{\hbox{\footnotesize th-nucl}} \approx 0.2\,\frac{M_{\rm III}}{m_N}
=2.4\times10^{56}~\frac{M_{\rm III}}{{\rm M}_\odot}\,,
\end{equation}
with less than a factor of 2 uncertainty caused by the stellar mass.
Notice that when using this expression in Eq.~(\ref{flux2}) we find
that the total flux does not depend on the chosen value for $M_{\rm
III}$. Even assuming a more realistic IMF one expects only a weak
dependence on it.

\begin{figure}[!ht]
\bigskip
\begin{center}
\epsfig{file=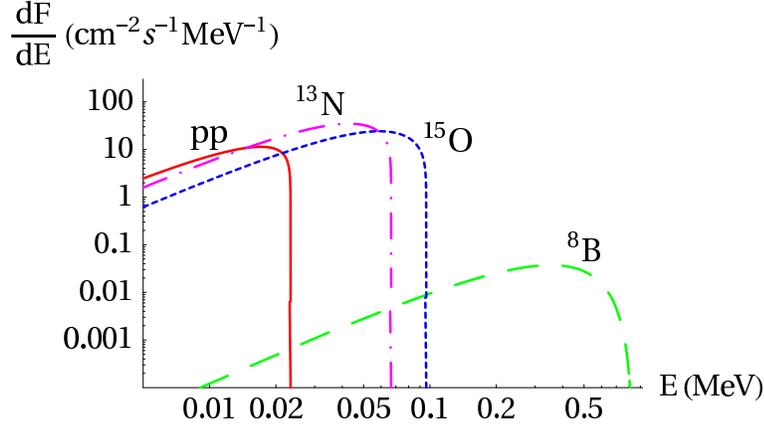,width=10truecm}
\caption{Diffuse cosmic neutrino flux from thermonuclear reactions in
  PopIII stars under the assumptions outlined in the text and a
  star-formation rate peaked at $\bar z=17$.
\label{fig:thermonuclear}}
\end{center}
\bigskip
\end{figure}

The energy spectrum depends on the burning mechanism. As the PopIII
stars are metal-free, one would expect a dominance of the pp chains
that does not require CNO elements. In reality, after the very initial
stages when the star produces its own metals through the
triple-$\alpha$ process, the energy production of a PopIII star is
dominated by the CNO burning of H into
He~\cite{Ezer71,Baraffe:2000dp,Marigo:2002nf}. From the tables in
Ref.~\cite{Ezer71}, for a very massive star the pp cycle is not
expected to contribute more than few percent at the central core,
while its importance possibly grows towards external layers. For
illustrative purposes we calculate the emission spectra under the
simplified hypothesis that 90\% of the neutrino flux is produced
through the CNO cycle and the remaining 10\% by the pp chains.  We
limit our considerations to the main source of the pp-I and pp-III
chains and the two leading channels of the main CNO cycle, i.e.
\begin{eqnarray}
\hbox to5.5cm{$p+p\rightarrow {}^{2}{\rm H}+e^{+}+\nu_e$\hfil}&
\hbox to5.5cm{Endpoint${}\approx0.42$~MeV\hfil}\nonumber\\
\hbox to5cm{${}^8{\rm B}\rightarrow{}^4{\rm He}+{}^4{\rm He}+
e^{+}+\nu_e$\hfil}
&\hbox to5.5cm{Endpoint${}\approx15$~MeV\hfil}\nonumber\\
\hbox to5cm{${}^{13}{\rm N}\rightarrow {}^{13}{\rm C}
+e^{+}+\nu_e$\hfil}
&\hbox to5.5cm{Endpoint${}\approx1.20$~MeV\hfil}\nonumber\\
\hbox to5cm{${}^{15}{\rm O}\rightarrow {}^{15}{\rm N}
+e^{+}+\nu_e$\hfil}
&\hbox to5.5cm{Endpoint${}\approx1.73$~MeV\hfil}
\end{eqnarray}
The standard theory of nuclear weak processes~\cite{nuastro} was used
to calculate the corresponding fluxes which are shown in
Fig.~\ref{fig:thermonuclear}.

A brief comment is in order.  In our Sun, approximately 85\% of
${}^3{\rm He}$ burns to ${}^4{\rm He}$ by the reaction
\begin{equation}\label{2he3}
{}^3{\rm He}+{}^3{\rm He}\rightarrow 2p+{}^4{\rm He}
\end{equation}
while the remaining 15\% goes into ${}^7{\rm Be}$ through
\begin{equation}\label{he3he4}
{}^4{\rm He}+{}^3{\rm He}\rightarrow {}^7{\rm Be}+\gamma
\end{equation}
finally forming ${}^7{\rm Li}$ via electron capture.  This secondary
path (pp-II chain) is responsible for the $^7{\rm Be}$ neutrinos,
while only a small fraction of about 0.02\% of beryllium further burns
to boron via the proton capture
\begin{equation}\label{b8}
{}^7{\rm Be}+p\rightarrow {}^8{\rm B}+\gamma\,.
\end{equation}
This situation changes somewhat in hotter stars like the ones we are
considering.  They spend their main-sequence life at $T_{\rm core}\sim
10^8$~K or larger so that the process Eq.~(\ref{b8}) is far less
Coulomb suppressed, allowing the boron channel to dominate over the
beryllium electron capture.  Further, a higher fraction of ${}^3{\rm
He}$ burns radiatively through the process Eq.~(\ref{he3he4}) with
respect to the solar case, dominated by the channel Eq.~(\ref{2he3}).
This suggests that the weight of the boron neutrinos relative to the
pp and Be ones significantly increases.  A detailed analysis would be
required to produce firm quantitative predictions, but from an
inspection of the rates it appears not unlikely that the boron flux is
as large as 10\% of the pp flux, a ratio that we will use in the
following.  The beryllium flux is marginal and will be neglected,
especially because of the leading role of the CNO reactions in the
same energy window.

For the flux prediction in Fig.~\ref{fig:thermonuclear} we have assumed 
a star formation rate peaked at a redshift $\bar z=17$.  We observe that
varying the value of $\bar{z}$ in the range 11--23 produces a change
of the amplitude and endpoint of the neutrino spectrum by a factor two
only.  For our purposes, the instantaneous reionization approximation
is then well justified.

A more careful estimate requires a detailed evolutionary model for the
PopIII generation, but our result already shows an interesting
feature. If one compares the flux of Fig.~\ref{fig:thermonuclear} with
the thermonuclear neutrino background from later star generations
shown in Ref.~\cite{Porciani:2003zq}, the PopIII background in the
approximate range of 10--100~keV is likely to be comparable to the
fluxes from ordinary stars in the entire universe and the galaxy,
except along the galactic disk.  As the flux from galactic and
extra-galactic plasma and cooling neutrinos in that range is expected
at most of the same order~\cite{Brocato:1997tu} we see that there
could be a region where the contribution of the PopIII stars to the
neutrino background is significant. Moreover, the spectral features of
this flux would give important information on key parameters such as
the reionization epoch, which can be determined by the energy
endpoint, and the fraction $f_{\rm III}$ which sets the integrated
flux.

Unfortunately, the detection of such a low-energy flux seems almost
impossible at present, in particular since it is overwhelmed by the
solar neutrino flux by orders of magnitude.  In the energy range up to
a few MeV, existing detection techniques are based on radiochemical
processes which do not provide directional information that would
allow one to discriminate against solar neutrinos.  Still, according
to our estimate the thermonuclear neutrino flux from PopIII stars is a
significant contribution to the diffuse cosmic neutrino background in
the 10--100~keV energy range.

The thermonuclear neutrinos are produced in the $\nu_e$ flavor state.
The well-established evidence for neutrino flavor
mixing~\cite{Maltoni:2004ei} implies that flavor transformations will
occur on the way to us.  In particular, the low energy part of the
spectrum will be mainly affected by vacuum oscillations (pp and CNO)
as in the sun, while matter effects will be dominating the oscillation
phenomena at higher energies leading to an adiabatic $\nu_e$
conversion. This means that the original $\nu_e$ in this case leave
the star in the $\nu_1$ mass eigenstate that contains significant
admixture of all three flavors $\nu_e$, $\nu_\mu$ and
$\nu_\tau$. Therefore, the thermonuclear neutrino flux arriving at
Earth will contain comparable contributions from all flavors.  A
detailed study of the oscillation phenomenon in PopIII stars is beyond
the aim of this paper as it requires detailed information of the
stellar density profiles as well as of the redshift dependence of the
PopIII star-formation rate.

%%%%%%%%%%%%%%%%%%%%%%%%%%%%%%%%%%%%%%%%%%%%%%%%%%%%%%%%%%%%%%%%%%%%%%
\section{Thermal neutrinos}
%%%%%%%%%%%%%%%%%%%%%%%%%%%%%%%%%%%%%%%%%%%%%%%%%%%%%%%%%%%%%%%%%%%%%%

\label{sec:thermal}

When a massive star ignites the carbon cycle, neutrino emission starts
to play a primary role in carrying the energy away from the core.  The
energy-loss rate depends sensitively on temperature so that detailed
knowledge of the late evolutionary stages is crucial to derive
reliable estimates of the neutrino fluxes.  However, the simulation of
these advanced stages is tricky. The silicon-burning stage constitutes
a ``potentially numerical unstable stage'' of stellar
modeling~\cite{WHW02}, especially because of its coupling to
convection phenomena.  Moreover, the burning phase of very massive
stars would be far from a quasi-stationary path, being eventually
characterized by a pair-instability phase and violent mass-ejecting
pulsations~\cite{FWH00,WHW02}.  Keeping in mind this caveat, one may
conservatively estimate the number of these thermal neutrinos as
follows.

Assume that the thermal neutrinos carry away all of the nuclear energy
of the star that is produced by carbon burning and all subsequent
burning phases.  From Fig.~1 of Ref.~\cite{Heger:2001cd} one deduces
that even for high progenitor masses (i.e.~200--300 M$_{\odot}$) only
up to about $30\,M_{\odot}$ burn to $^{28}$Si or heavier nuclei in the
late evolutionary phases.  Therefore, assuming that 15\% of the baryon
matter is involved in these advanced burning phases, under the extreme
hypothesis that all these nucleons are converted from carbon to
iron-group nuclei, and neglecting the additional energy supplied by
gravitational contraction, one obtains
\begin{equation}
N_{\rm thermal}\approx 0.15\,\frac{M_{\rm III}}{m_N}\,\,
\frac{B_{\rm Fe}-B_{\rm C}}{\langle E_{\nu}\rangle}\,, 
\end{equation}
where $B_{\rm Fe}\approx 8.8$~MeV is the binding energy per nucleon of
the iron-group nuclei, $B_{\rm C}\approx7.7$~MeV that of carbon, and
$\langle E_{\nu}\rangle$ is the average energy of the emitted
neutrinos.  For carbon burning and beyond, the thermonuclear reactions
take place at temperatures in the range 50--300~keV where the neutrino
losses are dominated by the pair-annihilation process
$e^{+}+e^{-}\rightarrow \nu+\bar{\nu}$ with neutrino energies in a
relatively narrow range of about
0.5--3~MeV~\cite{Shi:1998nd,Odrzywolek:2003vn}.  Taking $\langle
E_\nu\rangle=1.5$~MeV we thus find
\begin{equation}\label{eq:nthermal}
N_{\rm thermal}\approx
1.3\times10^{56}~\frac{M_{\rm III}}{{\rm M}_\odot}\,.
\end{equation}
A comparison with the other neutrino sources is shown in
Fig.~\ref{fig:fig5}, where a blackbody distribution with $T=0.5$~MeV
was assumed and the total flux was normalized to
Eq.~(\ref{eq:nthermal}).

Given the existing uncertainties, the best one can say is that this is
likely to be a significant contribution to the overall neutrino
spectrum from a PopIII star, but also quite dependent on its
evolutionary details.  As the redshift would move the energies into
the keV to few hundreds of keV range, once again this flux would be
dominated at low energies by the solar thermal neutrinos and
anti-neutrinos~\cite{Haxton:2000xb}, and at higher energies by the
thermonuclear neutrinos from the Sun ($\nu$) or by geo-neutrinos
($\bar{\nu}$), thus preventing any possible detection.  By comparing
our results with the curves shown in Ref.~\cite{Porciani:2003zq} one
expects a sub-leading contribution to the diffuse cosmic neutrino
background in this energy window.  Incidentally, we note that a
detailed evaluation of the thermal neutrino flux from galactic and
extragalactic normal stars (PopI and PopII) is still missing.

The thermal neutrinos are produced in all flavor states, although the
$\nu_e\bar\nu_e$ channel dominates. Flavor oscillations will lead to
partial swapping of the different fluxes. In any event, the fluxes
arriving at Earth will contain comparable contributions from neutrino
and anti-neutrinos of all flavors.

%%%%%%%%%%%%%%%%%%%%%%%%%%%%%%%%%%%%%%%%%%%%%%%%%%%%%%%%%%%%%%%%%%%%%%
\section{Core collapse}
%%%%%%%%%%%%%%%%%%%%%%%%%%%%%%%%%%%%%%%%%%%%%%%%%%%%%%%%%%%%%%%%%%%%%%

\label{sec:corecollapse}

We assume that every PopIII star ends its life in a core collapse (CC)
that produces a total number of $\nu_\alpha$ given by
\begin{equation}\label{NSN}
N^\alpha_{\rm CC}= \frac{E^\alpha_{\rm tot}}{\langle
E_{\alpha}\rangle}\,. 
\end{equation}
Here, $E_{\rm tot}^{\alpha}$ is the total energy released in the
neutrino species $\alpha=\nu_e$, $\nu_\mu$, $\nu_\tau$, $\bar\nu_e$,
$\bar\nu_\mu$, and $\bar\nu_\tau$, whereas $\langle E_{\alpha}\rangle$
is the corresponding average neutrino energy.  Our following estimates
depend primarily on the rotating 300 M$_{\odot}$ simulation performed
in Ref.~\cite{FWH00} that to our knowledge is the most accurate
analysis of its kind.

\begin{figure}[!ht]
\bigskip
\begin{center}
\epsfig{file=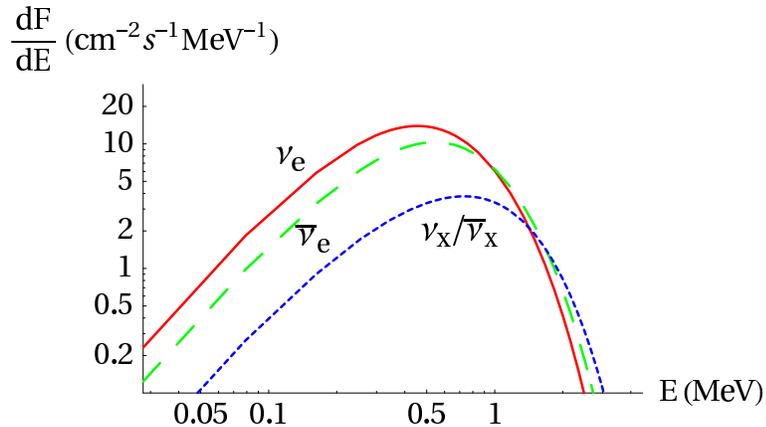,width=10truecm}
\caption{Diffuse cosmic neutrino fluxes from core collapse of PopIII
  stars. The $\nu_e$ (solid line), $\bar{\nu}_e$ (long-dashed) and
  $\nu_x$ (short-dashed) spectra are shown for a star-formation rate
  peaked at $\bar z=17$ and for the fluxes calculated as explained in
  the text.} \label{fig:fig1}
\end{center}
\bigskip
\end{figure}

The total energy release $E_{\rm tot}^{\alpha}$ is of the order of
$10^{55}$~erg for each neutrino species $\alpha$ as indicated by
Fig.~9 of Ref.~\cite{FWH00} that shows a luminosity for each neutrino
species in the range $10^{54}$--$10^{55}~{\rm erg}~{\rm s}^{-1}$.  An
almost constant luminosity lasts for a few seconds until the event
horizon reaches the neutrino sphere, eventually cutting off the
neutrino emission.  During the luminosity peak the equipartition of
energy among different neutrino species holds approximately and the
mean energies satisfy a hierarchical pattern similar to standard
core-collapse supernovae~\cite{Keil:2002in}.  At the peak of the
luminosity curve, the star is optically thick and different neutrino
species are trapped inside the core at different radii because of the
different trapping reactions. Denoting with ${\nu}_x$ the ${\mu}$ and
${\tau}$ neutrinos and antineutrinos, the average energies are
numerically found to be~\cite{FryerPC}
\begin{equation}\label{enpeak}
\langle E_{\nu_e}\rangle\approx15~{\rm MeV}\,,
\quad
\langle E_{\bar{\nu}_e}\rangle\approx17~{\rm MeV}\,,
\quad
\langle E_{\nu_x}\rangle \approx20~{\rm MeV}\,.
\end{equation}
After the neutrino spheres have disappeared in the black hole, 
smaller contributions to the luminosity are still provided by
neutrinos escaping from outer layers of the star with slightly lower
average energies~\cite{FryerPC},
\begin{equation}\label{enpeak2}
\langle E_{\nu_e}\rangle\approx 10~{\rm MeV}\,,
\quad
\langle E_{\bar{\nu}_e}\rangle\approx 12~{\rm MeV}\,,
\quad
\langle E_{\nu_x}\rangle \approx 9~{\rm MeV}\,.
\end{equation}
The neutrino spectra as seen by a distant observer can be approximated
in terms of a normalized Fermi-Dirac distribution
\begin{equation}\label{fermidirac}
\frac{dN^\alpha_i(M_{\rm III},E(1+z))}{dE} = \frac{2}{3 \zeta_3
T_\alpha^3} \, \frac{E^2
(1+z)^2}{e^{E(1+z)/T_\alpha}+1}
\end{equation} 
where $\langle E_{\alpha}\rangle=7\pi^4\,T_\alpha/(180\zeta_3)$
defines the effective temperature $T_\alpha$. The temperature adopted
in this case is the properly averaged mean between the typical values
of the two different luminosity phases.  Our results for the neutrino
fluxes are shown in Fig.~\ref{fig:fig1} assuming a star-formation rate
$\rho(z)$ peaked at $\bar z=17$. Obviously, the inclusion of neutrino
oscillations would smooth the flavor-dependent flux differences shown
in Fig.~\ref{fig:fig1}.

\begin{figure}[!ht]
\bigskip
\begin{center}
\epsfig{file=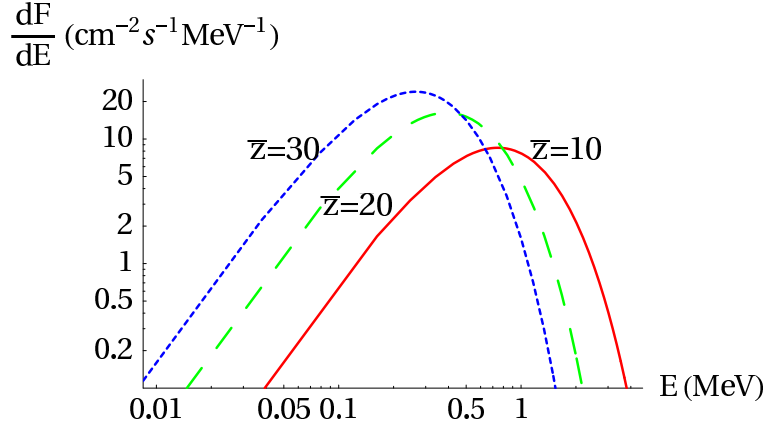,width=10truecm}
\caption{Diffuse cosmic $\nu_e$ flux from PopIII core collapse for an
  emission redshift $\bar z=10$ (solid line), $\bar z=20$
  (long-dashed) and $\bar z=30$ (short-dashed).}
\label{fig:fig2}
\end{center}
\bigskip
\end{figure}

\begin{figure}[!hbtp]
\bigskip
\begin{center}
\epsfig{file=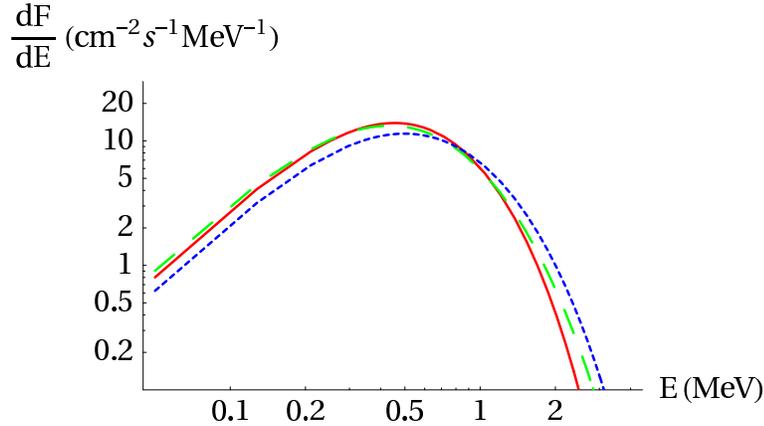,width=10truecm}
\caption{Diffuse cosmic $\nu_e$ flux from PopIII core collapse for an
  average redshift $\bar{z}=17$.  The solid line is for a
  star-formation rate $\rho(z)=\delta(z-\bar z)$, while the
  long-dashed and short-dashed curves correspond to a constant rate as
  a function of redshift and time, respectively, during a redshift
  interval $z_i=24$ and $z_f=10$.} \label{fig:fig3}
\end{center}
\bigskip
\end{figure}

The effect of the cosmic redshift, as expected, is quite dramatic as
can be seen from Fig.~\ref{fig:fig2}, where we show the $\nu_e$
spectrum for three different values $\bar z=10$, 20 and 30. On the
other hand, the flux is much less sensitive to the redshift dependence
of the star-formation rate $\rho(z)$. This can be seen from
Fig.~\ref{fig:fig3}, where we show the $\nu_e$ flux for a delta-like
$\rho(z)$, along with the results for a constant star-formation rate
as a function of redshift or time. We have fixed in the first case
$\bar{z}=17$, while in the latter ones we have chosen the generous
redshift interval $z_i=24$ and $z_f=10$. The spreading of PopIII
formation over a non-zero redshift interval has the main effect of
slightly shifting the spectrum towards higher energies, where however
the flux becomes very small, by keeping constant the integrated flux.

These results were obtained assuming a simple IMF, peaked at $300\,
{\rm M}_\odot$.  However, changing the stellar mass distribution
affects the neutrino flux from core collapse events because the amount
of neutrino energy per baryon that is released in a core collapse
event depends on the stellar mass.  We recall that for a progenitor of
$20\,{\rm M}_\odot$ the core collapse leading to an ordinary supernova
produces a total neutrino energy release of about $3\times10^{53}$~erg
\cite{Keil:2002in}, two orders of magnitude smaller than what is found
in Ref.~\cite{FWH00} for a rotating $300\,{\rm M}_\odot$ star that has
about ten times the mass. Therefore, in the latter case the neutrino
energy per baryon that is released is about ten times larger.  Also,
the average neutrino energies would be different for different stellar
masses. Finally, different choices of stellar parameters could
dramatically reduce the neutrino flux escaping from the star before
the collapse, as shown in Ref.~\cite{FWH00} for the non-rotating star
simulation.  The neutrino fluxes from PopIII core collapse shown here
are therefore somewhat of an upper limit and, for a given fraction of
baryons $f_{\rm III}$ forming PopIII stars, the true diffuse cosmic
flux could be smaller by at least a factor of a few.

The detection possibility of the PopIII neutrino flux from the core
collapse mechanism is again extremely challenging because the peak of
the energy distributions likely falls in the solar pp range.  Even for
more optimistic model parameters that shift the peak of the
distribution to energies beyond $0.42$~MeV, the signal still would be
several orders of magnitude smaller than the sub-dominant CNO solar
flux.  In the anti-neutrino channel, the cosmic flux is overwhelmed by
the geo- and reactor-neutrinos.  Geo-neutrinos have a flux in the
energy window up to about 3~MeV of $10^{4}$--$10^{5}~{\rm
cm}^{-2}~{\rm s}^{-1}~{\rm MeV}^{-1}$ \cite{Krauss:1983zn}.

%%%%%%%%%%%%%%%%%%%%%%%%%%%%%%%%%%%%%%%%%%%%%%%%%%%%%%%%%%%%%%%%%%%%%%
\section{Conclusions}
%%%%%%%%%%%%%%%%%%%%%%%%%%%%%%%%%%%%%%%%%%%%%%%%%%%%%%%%%%%%%%%%%%%%%%

\label{sec:conclusions}

\begin{figure}[!ht]
\bigskip
\begin{center}
\epsfig{file=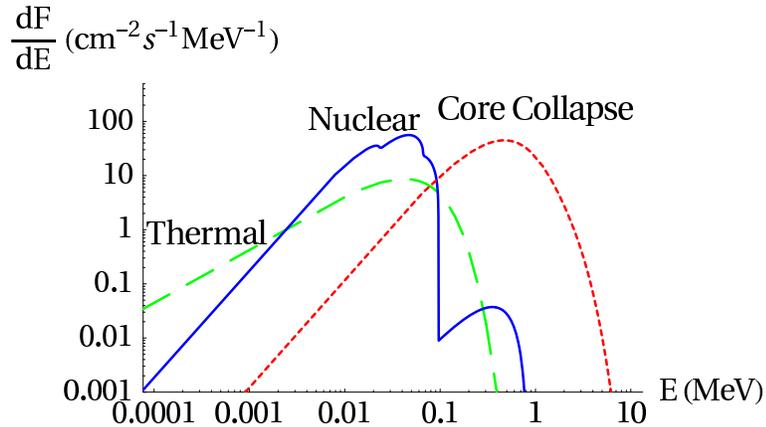,width=10truecm}
\caption{Diffuse cosmic neutrino flux from PopIII stars caused by
thermal emission (long-dashed line), thermonuclear reactions (solid
line), and core collapse (short-dashed line), assuming a
star-formation rate peaked at $\bar z=17$.  All of these flux
estimates should be taken as optimistic---the true fluxes could be
lower by factors of a few or even an order of magnitude, especially
for the CC case. The thermonuclear flux consists of neutrinos of all
flavors due to oscillations. The thermal and core-collapse fluxes
contain both neutrinos and anti-neutrinos of all flavors.}
\label{fig:fig5}
\end{center}
\bigskip
\end{figure}

We have estimated the diffuse cosmic neutrino flux produced by the
pregalactic generation of stars known as Population III.  Using very
simple models, we have calculated the fluxes due to the thermonuclear
hydrogen burning processes, the contribution of the thermal neutrinos
emitted in the late evolutionary stages, and the neutrino burst
emitted during the core collapse phase at the end of the stellar
evolution process. We summarize our optimistic estimates for these
fluxes in Fig.~\ref{fig:fig5}. We find that these fluxes are too small
to be detected with present or planned experiments: one expects indeed
less than $10^{-4}\,{\rm yr}^{-1}$ events per kton in a typical water
Cherenkov detector. Moreover, even our optimistic estimates imply that
local backgrounds such as solar or geo-neutrinos completely dominate
the cosmic diffuse flux.  The only speculative neutrino detection
channel for PopIII stars thus remains that of the ultra-high energy
neutrinos emitted in a GRB-like process~\cite{Schneider:2002sy}.

Turning this result around, we conclude that even under extreme
assumptions the diffuse cosmic neutrino flux from PopIII stars is
irrelevant as a possible contamination or background to existing
neutrino experiments.

\begin{figure}[!ht]
\bigskip
\begin{center}
\epsfig{file=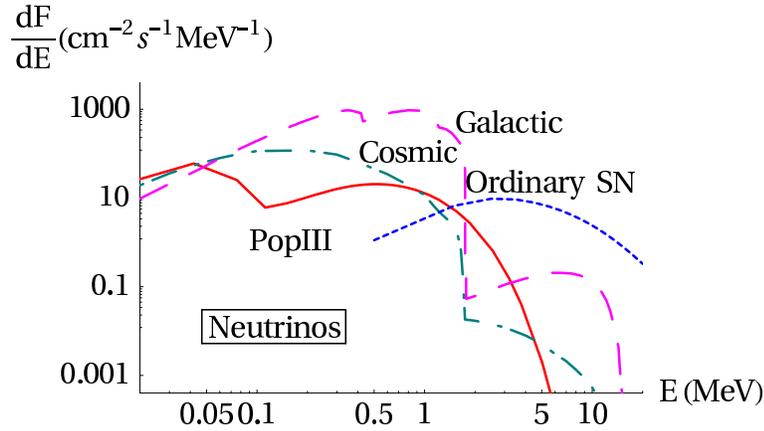,width=10truecm}
\caption{Diffuse cosmic neutrino flux from PopIII stars (solid line),
summed over all production mechanisms. For comparison we show the
diffuse cosmic and galactic flux from thermonuclear reactions in
ordinary stars according to Ref.~\cite{Porciani:2003zq} and from
ordinary core-collapse supernovae.}
\label{fig:fig7}
\end{center}
\bigskip
\end{figure}

\begin{figure}[!ht]
\bigskip
\begin{center}
\epsfig{file=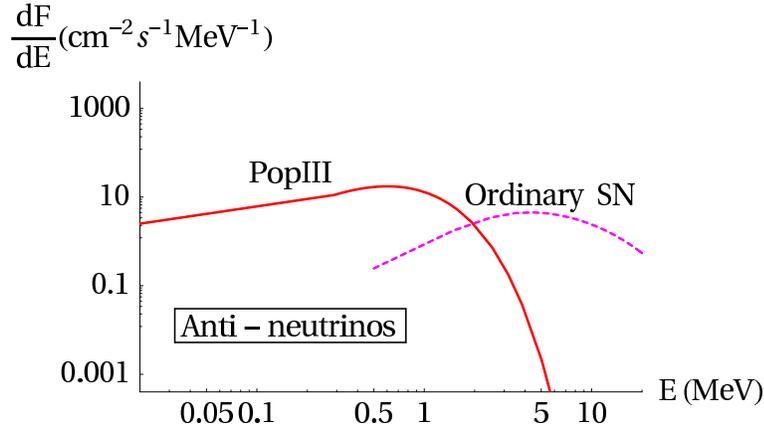,width=10truecm}
\caption{Diffuse cosmic anti-neutrino flux from PopIII stars (solid
line), summed over all production mechanisms. For comparison we show
the diffuse flux from ordinary core-collapse supernovae.}
\label{fig:fig8}
\end{center}
\bigskip
\end{figure}

While perhaps an academic exercise in view of realistic detection
possibilities, we find it an interesting question in its own right to
develop a quantitative understanding of the diffuse cosmic neutrino
fluxes due to all astrophysical sources and in all energy bands. In
principle, they could be disentangled from the local contributions
(Sun, Earth) thanks to directional information of the latter ones,
provided that the instrumental noise is sufficiently under control and
sufficiently long integration times are allowed.  The diffuse cosmic
and galactic neutrino flux from ordinary stars was previously
estimated~\cite{Porciani:2003zq}. We compare our estimates for the
PopIII flux with the most optimistic predictions of
Ref.~\cite{Porciani:2003zq} in our Fig.~\ref{fig:fig7}.  For a baryon
fraction $f_{\rm III}=10^{-3}$ forming PopIII stars as assumed in all
of our estimates, we find the two fluxes to be comparable, at least
within the current uncertainties.  (Of course, the galactic flux is
strongly peaked along the galactic disk and about 2~orders of
magnitude weaker in the orthogonal direction so that it does not
directly compare with the isotropic cosmic flux.)  Notice that at low
energies the flux estimate from ordinary stars does not include
neutrinos from thermal plasma reactions.  Remarkably, it is likely
that the $\bar{\nu}$ flux from PopIII is the dominant cosmic component
in the 0.01--few MeV range~(Fig.~\ref{fig:fig8}).

The one diffuse cosmic neutrino flux that may become detectable within
the next few years is that from all ordinary core-collapse supernovae
in the universe. In the energy range of a few tens of MeV, the
expected flux exceeds all
backgrounds~\cite{Strigari:2003ig,Ando:2004hc} and may become
detectable in the possible future GADZOOKS
version~\cite{Beacom:2003nk} of Super-Kamiokande, in a corresponding
megaton detector, or in a 100~kton liquid Argon
experiment~\cite{Cocco:2004ac}.  For comparison we show this diffuse
flux also in Fig.~\ref{fig:fig7}, calculated according to the
``median'' model described in Ref.~\cite{Strigari:2003ig,Cocco:2004ac}.  
Although the low-energy tail is affected by large uncertainties because of the
significant role of the early (i.e.~high redshift) SN contribution to
the rate, it is clearly the dominant diffuse cosmic neutrino and
anti-neutrino flux in the energy range above a few~MeV as shown in
Figs.~\ref{fig:fig7} and~\ref{fig:fig8}.

%\newpage

%%%%%%%%%%%%%%%%%%%%%%%%%%%%%%%%%%%%%%%%%%%%%%%%%%%%%%%%%%%%%%%%%%%%%%
\ack
%%%%%%%%%%%%%%%%%%%%%%%%%%%%%%%%%%%%%%%%%%%%%%%%%%%%%%%%%%%%%%%%%%%%%%

We thank the organizers and sponsors of the workshop ``Chemical
Enrichment of the Early Universe,'' (Santa Fe, New Mexico, August
2004) for stimulating discussions. In particular F.~I.~warmly thanks
T.~Abel for useful discussions and C.~L.~Fryer for providing details
of the 300~M$_\odot$ star numerical simulation.  The authors also
aknowledge V. Pettorino and C. Kiessig for fruitful discussions on
ordinary supernovae diffuse fluxes.  In Munich, this work was partly
supported by the Deut\-sche For\-schungs\-ge\-mein\-schaft under grant
SFB 375 and the ESF network Neutrino Astrophysics.

%%%%%%%%%%%%%%%%%%%%%%%%%%%%%%%%%%%%%%%%%%%%%%%%%%%%%%%%%%%%%%%%%%%%%%
% References
%%%%%%%%%%%%%%%%%%%%%%%%%%%%%%%%%%%%%%%%%%%%%%%%%%%%%%%%%%%%%%%%%%%%%%

\end{document}